\documentclass[amsmath,amssymb]{nature}
\usepackage{bm}
\usepackage{color}
\usepackage{gensymb}
\newcommand{\eeq}{\end{equation}}

\usepackage[dvipdfmx]{graphicx}

\title{Large anomalous Nernst effect at room temperature in a chiral antiferromagnet} 

\author{Muhammad Ikhlas$^1$\footnote[1]{These two authors contributed equally.}, Takahiro Tomita$^1$\footnotemark[1], Takashi Koretsune$^{2,3}$, Michi-To Suzuki$^{2}$, Daisuke Nishio-Hamane$^1$, Ryotaro Arita$^{2,4}$, Yoshichika Otani$^{1,2,4}$, Satoru Nakatsuji$^{1,4,*}$\\}

\begin{document}
	
	\maketitle
	
	\begin{affiliations}
		\item Institute for Solid State Physics, University of Tokyo, Kashiwa 277-8581, Japan
		\item RIKEN-CEMS, 2-1 Hirosawa, Wako 351-0198, Japan		
		\item PRESTO, Japan Science and Technology Agency (JST), 4-1-8 Honcho Kawaguchi, Saitama 332-0012, Japan.
		\item CREST, Japan Science and Technology Agency (JST), 4-1-8 Honcho Kawaguchi, Saitama 332-0012, Japan.

	\end{affiliations}

\begin{abstract}
Temperature gradient in a ferromagnetic conductor may generate a spontaneous transverse voltage drop in the direction perpendicular to both magnetization and heat current. This anomalous Nernst effect (ANE) has been considered to be proportional to the magnetization\cite{Smith1911,Kondorskii1964,Lee2004,Miyasato2007,Yong2008,Sakuraba2013,Hasegawa2015}, and thus observed only in ferromagnets, while recent theories indicate that  ANE provides a measure of the Berry curvature at the Fermi energy  $E_{\rm F}$\cite{Xiao2006,Niu2010}. Here we report the observation of a large ANE at zero field in the chiral antiferromagnet Mn$_3$Sn\cite{Mn3Sn}.  Despite a very small magnetization $\sim 0.002$ $\mu_{\rm B}/$Mn, the transverse Seebeck coefficient at zero field is $\sim 0.35~\mu$V/K at room temperature and reaches $\sim 0.6~\mu$V/K at 200 K, comparable with the maximum value known for a ferromagnetic metal.  Our first-principles calculation reveals that the large ANE comes from a significantly enhanced Berry curvature associated with the Weyl points nearby $E_{\rm F}$\cite{Yang2016}. 
%, and several orders of magnitude larger than the ANE expected based on the conventional scaling law with $\mbox{\boldmath $M$}$. 
%In addition, the sign of the Nernst signal can be flipped by rotating $\mbox{\boldmath $M$}$ with a small field of $\sim 100$ Oe. 
The ANE is geometrically convenient for the thermoelectric power generation, as it enables a lateral configuration of the modules to efficiently cover the heat source\cite{Sakuraba2013}. Our observation of the large ANE in an antiferromagnet paves a way to develop a new class of thermoelectric material using topological magnets to fabricate an efficient, densely integrated thermopile.
\end{abstract}

Current intensive studies on thermally induced electron transport in ferromagnetic (FM) materials have opened various venues for the research on thermoelectricity and its application\cite{Bauer2012,Uchida2008,Slachter2010,Huang2011}. This trend has also triggered renewed interest in anomalous Nernst effect (ANE) in FM metals\cite{Lee2004,Miyasato2007,Yong2008,Huang2011,Sakuraba2013,Hasegawa2015}, which is the spontaneous transverse voltage drop induced by heat current and is known to be proportional to magnetization (Fig. 1a). 
On the other hand, the recent Berry phase formulation of the transport properties has led to the discovery that a large anomalous Hall effect (AHE) may arise not only in ferromagnets, but in antiferromagnets and spin liquids, in which the magnetization is vanishingly small\cite{Nagaosa2010,Bruno2007,Shindou2001,Machida2010,Chen2014,Kubler2014,Mn3Sn,Kiyohara2016,Nayak2016}.
As the first case in antiferromagnets, Mn$_3$Sn has been experimentally found to exhibit a large AHE\cite{Mn3Sn}.  While the AHE is obtained by an integration of the Berry curvature for all the occupied bands, the ANE is determined by the Berry curvature at $E_{\rm F}$\cite{Xiao2006,Niu2010}. Thus, the observation of a large AHE does not guarantee the observation of a large ANE. Furthermore, the ANE measurement should be highly useful to clarify the Berry curvature spectra near $E_{\rm F}$ and to verify the possibility of the Weyl metal recently proposed for Mn$_3$Sn\cite{Yang2016}. 

Mn$_3$Sn has a hexagonal crystal structure with space group of $P6_3/mmc$\cite{Kren1975}. Mn atoms form a breathing type of a kagome lattice in the $ab$-plane (Fig. 1b), and the Mn triangles constituting the kagome lattice are stacked on top along the $c$-axis forming a tube of face sharing octahedra. On cooling below the N\'{e}el temperature of 430 K, Mn magnetic moments of $\sim 3 \mu_{\rm B}$ lying in the $ab$-plane form a coplanar, chiral magnetic structure characterized by $Q=0$ wave vector,  as clarified by the previous neutron diffraction studies\cite{TomiyoshiYamaguchi1982,Brown1990}.  The combination of geometrical frustration and Dzyaloshinskii$-$Moriya interaction leads to the inverse triangular spin structure with uniform vector chirality (Fig. 1b)\cite{TomiyoshiYamaguchi1982,Brown1990,Nagamiya1982}.  The chiral antiferromagnetic order has orthorhombic symmetry and thus induces a very tiny magnetization $\sim 2$ m$\mu_{\rm B}$/Mn, which is essential to switch the non-collinear antiferromagnetic structure by using magnetic field.   On further cooling below $\sim 50$ K, a cluster glass phase appears as spins cant toward [0001] ($c$ axis)\cite{Tomiyoshi1986,Brown1990,Feng2006}.  In this study, we used two as-grown single crystals that have a single phase of hexagonal Mn$_3$Sn \cite{Ohmori1987} (Methods, Supplementary Information), with slightly different compositions, i.e., Mn$_{3.06}$Sn$_{0.94}$ for ``Sample 1'', and  Mn$_{3.09}$Sn$_{0.91}$ for ``Sample 2'' (Methods). Hereafter, we focus on the coplanar magnetic phase at $T >$ 60 K. 

We first provide clear evidence of the large anomalous Hall and Nernst effects observed in Samples 1\&2. Figures 2a \& 2b show the field dependence of the Hall resistivity $\rho_{\rm H}(B)$ in $B \parallel [01\bar{1}0]$ for Samples 1 \& 2, respectively. Clearly, there is a sharp jump in $\rho_{\rm H}(B)$ with a small coercivity of $< 200$ G. 
%In sharp contrast, no field dependence was found in the longitudinal resistivity (Supplementary Information). 
In particular for Sample 2, the size of the jump $\Delta\rho_{\rm H}$  reaches $\sim 9~\mu\Omega\mathrm{cm}$  at 100 K, which would be equivalent to the Hall resistivity due to an ordinary Hall effect under $\sim$ a few 100 T for free conduction electrons with density of order one electron per Mn atom. To make comparison with theory later, here we take the $x$, $y$, and $z$ coordinates along $[2\bar{1}\bar{1}0]$, $[01\bar{1}0]$, and $[0001]$, and estimate the Hall conductivity employing the following expression that takes care of the anisotropy of longitudinal resistivity (Fig. S1), namely, $\sigma_{ji} \approx -\rho_{ji}/(\rho_{jj}\rho_{ii})$, where $(i,j)=(x,y),\  (y,z),\ (z,x)$ \cite{Kiyohara2016} (Supplementary Information). Figures 2c \& 2d show $-\sigma_{zx}$ vs. $B$ for Samples 1 \& 2, respectively. The zero field value is $\sim 50~\Omega^{-1}\mathrm{cm}^{-1}$ at 300 K for both samples and it becomes particularly enhanced for Sample 2 at low $T$s and reaches 120  $\Omega^{-1}\mathrm{cm}^{-1}$ at 100 K. 
%While the size of the Hall resistivity is almost the same as the previously observed values\cite{Mn3Sn}, the coercivity is several times smaller, indicating less number of defects and/or impurities in the crystal used in the present study.  A smaller value of the longitudinal resistivity $\rho$ than the previous case was obtained (Fig. S1, Supplementary Information) and leads to the larger Hall conductivity $\sigma_{\rm H} \approx - \rho_{\rm H}/\rho^2$ as shown in Fig. 2b; it is 100 $\Omega^{-1}\mathrm{cm}^{-1}$ at 300 K and reaches 280 $\Omega^{-1}\mathrm{cm}^{-1}$ at 50 K. This is strikingly large, given the very small magnetization, and the enhancement in the Hall conductivity in comparison with the previous study \cite{Mn3Sn} may possibly come from the shift in the Fermi level due to the change in the Mn concentration (Supplementary Information, Fig. S4).  
The sign change in the Hall effect as a function of field should come from the flipping of the tiny uncompensated moment which follows the rotation of the sublattice moments\cite{Mn3Sn,Nagamiya1982,TomiyoshiYamaguchi1982}. %The small coercivity of $\sim100$ Oe indicates that the sublattice moment can be flipped almost freely, consistent with the theoretical calculation showing that the anisotropy energy is negligible up to the 4th order term\cite{Nagamiya1982,TomiyoshiYamaguchi1982}. 

Our main experimental observation of a large ANE at room temperature is provided in Fig. 3a. The Nernst signal (transverse thermopower) $-S_{zx}$ of Sample 1 shows a clear rectangular hysteresis with an overall change of $\Delta S_{zx} \sim 0.7 ~\mu$V/K as a function of the in-plane field. This is significantly large for an antiferromagnet and comparable to the values reported for ferromagnets, as we will discuss. While the in-plane Nernst signal exhibits hysteresis with almost no anisotropy, the out-of-plane $c$-axis component is zero within experimental accuracy, indicating no spontaneous effect in this direction. 
To further characterize the ANE, we compare it with the magnetization $M$ by plotting in Fig. 3a both $-S_{zx}$ and $M$ sharing an $x$-axis for the in-plane field. In low fields, the hystereses in both data almost overlap on top of each other. On the other hand, in the higher field region than the coercivity of $\sim 100$ G, the Nernst effect remains nearly constant, while $M$ increases linearly with field as the sub-lattice moments cant toward the external field direction. These indicate negligible contributions from the normal Nernst effect and presumably the conventional ANE. Furthermore, the remnant Nernst signal at zero field is nearly the same as the saturated value in high field, demonstrating that the single-domain crystal has a large spontaneous Nernst signal as the first case for an antiferromagnet. Similar behavior was found in the field cycles made for Sample 2 (Fig. 3b). A systematic change of the Nernst signal was observed with varying $T$ for both Samples 1 \& 2, (Figs. 3c \& 3d). For Sample 1, the Nernst signal peaks at $\sim 200$ K, reaching a value of 0.6 $\mu$V/K, comparable with the maximum value known for a FM metal at room temperature\cite{Hasegawa2015}, while Sample 2 shows a smaller value, peaking at 250 K with 0.3 $\mu$V/K. A distinct mechanism from the FM case should be at work here as the spontaneous $M$ in Mn$_3$Sn is nearly 1000 times smaller than ordinary ferromagnets.

To further characterize thermoelectric properties, we measured the longitudinal Seebeck coefficient $S_{ii}$ as a function of $T$ and $B$.  No field dependence was seen in $S_{ii}$ for both samples in the $T$ and $B$ range of the measurements (Supplementary Information, Figs. S2a\&b).
%, indicating that the subtraction of a zero-field value from raw data is sufficient to obtain a hysteresis curve of the Nernst signal. 
The Seebeck coefficient for Sample 1 positively peaks at $\sim 300$ K (Fig. 3c, inset), and takes a minimum value with a negative sign around 50 K. %On cooling, the Seebeck coefficient peaks again with a negative sign at around \color{red} 50 K, which coincides with the steep rise of heat capacity (Supplementary Information).  
This minimum coincides with the steep rise in the specific heat, and thus, this sign change in $S(T)$ may be related to the effect of phonon drag which typically occurs at $\sim \Theta_{\rm D}/5$, where $\Theta_{\rm D}$ is the Debye $T$ (Supplementary Information, {Fig. S3}). Much reduced or no low $T$ sign change was seen for Sample 2 (Fig. 3d, inset).%transition temperature of 50 K below which Mn moments start canting toward the $c$-axis\cite{Tomiyoshi1986,Brown1990,Feng2006}. 
%The thermoelectric Hall angle comparing the Nernst effect with the Seebeck effect increases on cooling, from 1 \% at 300 K to 10 \% at 100 K, which is significantly large and comparable with the observed value for ferromagnets. 

Generally, temperature gradient \mbox{\boldmath $\nabla T$} in an open circuit condition is known to produce longitudinal and transverse electric field \mbox{\boldmath $E$}, which can be expressed as $\mbox{\boldmath $J$} =  \mbox{\boldmath $\sigma \cdot E$} + \mbox{\boldmath $ \alpha   \cdot (- \nabla T)$}= 0$. Here \mbox{\boldmath $J$}, \mbox{\boldmath $\sigma$} and \mbox{\boldmath $\alpha$} are the current density, electrical conductivity tensor, and thermoelectric conductivity tensor, respectively. This contains the transverse electric field coming from the thermal Hall effect, which should be negligible as it is usually one order of magnitude smaller than other contributions\cite{Lee2004,Miyasato2007,hanasaki2008}. Assuming this, the Nernst signal can be expressed as $S_{ji} = \rho_{jj} (\alpha_{ji} - S_{ii} \sigma_{ji})$, using the Seebeck coefficient $S_{ii} (=S)$ and the Hall conductivity $\sigma_{ji} (=\sigma_{\rm H})$. For this analysis, we used the $T$ dependence of the Hall conductivity shown in Fig. 3e. As shown in Figs. 3c\&3d, $-\rho \alpha_{ji}$ estimated using this relation is found to be larger than $-S_{ji}$. ~We further estimated the transverse thermoelectric conductivity $-\alpha_{zx}$ vs. $T$ (Fig. 3f), which shows a systematic increase on cooling and has the maximum at $T \sim 150$(100) K for Sample 1(2). At low $T$s, Sample 2 has twice smaller values than Sample 1. In ferromagnets, $\alpha_{ji}$ normally shows an increase on cooling, scaling linearly with $M$, and decreases linearly with $T$ after $M$ saturates\cite{Lee2004,Miyasato2007}.  In Mn$_3$Sn, however, no correlation with $M$ was seen in $-\alpha_{zx}(T)$ (Fig. 3f). 
% at least down to 50 K. %This unusual behavior again points to a fully distinct mechanism for ANE.

To further demonstrate the qualitative difference between the ANE observed in Mn$_3$Sn and in ferromagnets, we made a full logarithmic plot of the anomalous Nernst signal vs. the magnetization for various FM metals and Mn$_3$Sn (Fig. 4, Methods). Here, the absolute values of the Nernst signal were taken in the magnetically ordered states of each material, and were plotted using $B$ and $T$ as implicit parameters. Similarly to AHE\cite{Nagaosa2010}, the ANE for ferromagnets is known to be proportional to magnetization. Indeed, Figure 4 confirms such an overall trend for a broad range of FM metals that the anomalous Nernst signal becomes larger with increasing magnetization (Methods). The shaded region which covers all the data points indicates that the anomalous Nernst signal is indeed roughly proportional to the magnetization $M$, i.e. $|S_{ji}| = |Q_{\rm s}|\mu_0M$ with the anomalous Nernst coefficient $|Q_{\rm s}|$ ranging between 0.05 and 1 $\mu$V/KT. Following this relation, Mn$_3$Sn would have produced the Nernst signal of the order of $0.01 \sim 2$ nV/K with the observed magnetization. Strikingly, however, $S_{ji}\sim 0.3 \mu$V/K found at room $T$ is more than 100 times larger than what would be expected based on the above scaling relation for ferromagnets. 
%This clearly indicates again a different type of mechanism for the Nernst effect in Mn$_3$Sn.

The significantly large anomalous Nernst and Hall effects in Mn$_3$Sn do not follow their conventional scaling relation with $M$, and thus it is natural to assume that both effects arise through the same mechanism distinct from the conventional one for ferromagnets. The inverse triangular spin structure for Mn$_3$Sn reduces the lattice symmetry from six fold to two fold in the plane and thus based on the symmetry argument, the Hall effect may appear in the $ab$-plane\cite{suzuki2016cluster}. Indeed, a recent calculation found that Mn$_3$Sn may have a large anomalous Hall conductivity\cite{Kubler2014}, which can be estimated by integrating the Berry curvature of the occupied bands over the entire Brillouin zone\cite{Niu2010}. More recently, the possibility of a Weyl metal has been proposed, where the bands crossing $E_{\rm F}$ have several Weyl points, around which the Berry curvature diverges\cite{Yang2016}.  
%Experimentally, we observed the Hall conductivity reaching $\sim 280~\Omega^{-1}\mathrm{cm}^{-1}$, which is consistent with the value estimated by the calculation\cite{Kubler2014}. 
%Theoretically, the anomalous Hall effect for a three dimensional material can be as large as $\sigma_{\rm H} =\frac{e^2}{2\pi h}|{\bf G}|$ as expected for the quantum Hall effect per atomic layer with the Chern number of unity, where  $e$, $h$ and $\mathbf{G}$ are the elementary charge and Planck constant, and the reciprocal lattice unit vector, respectively\cite{Nagaosa2010}. The observed value of $\sim 280~\Omega^{-1}\mathrm{cm}^{-1}$ is large and is of the same order of magnitude as $\sigma_{\rm H} =\frac{e^2}{2\pi h}|{\bf G}| \sim 800~\Omega^{-1}\mathrm{cm}^{-1}$ for the layered quantum Hall effect. This may be explained if the Weyl points are separated by a wave vector of $\sim2/5\mathbf{G}$. 

 Another intriguing quantity governed by the Berry curvature is the anomalous Nernst effect. While all the occupied bands are relevant for the anomalous Hall conductivity, only the Berry curvature around the Fermi level determines the ANE or, more precisely, the transverse thermoelectric conductivity $\alpha_{ji}$ \cite{Xiao2006,Niu2010}. Therefore, %as can be seen in the Mott relation $\alpha_{yx} = \frac{\pi^2 k_{\rm B}^2 T}{3 e} (\frac{\partial\sigma_{\rm H}}{\partial{E}})_{E_{F}}$, 
$\alpha_{ji}$ is significantly enhanced when the Berry curvature takes a large value at $E_{\rm F}$. To see this in the case of Mn$_3$Sn, we performed a first-principles calculation, confirming the Weyl points nearby $E_{\rm F}$\cite{Yang2016}. The calculated anomalous Hall conductivity $-\sigma_{zx}$ is found as large as seen in experiment.
%, and has a positive (negative) slope above (below) $E_{\rm F}$, forming a peak at the energy level where the Weyl points exist \cite{Yang2016} (Fig. S4). 
Theoretically, the extra Mn in Sample 1(2) should dope the conduction electron (Supplementary Information) and thus increase $E_{\rm F}$ by 0.04 (0.05) eV.  Since $-\sigma_{zx}$ forms a peak at $E-E_{\rm F} \sim 0.065$ eV (Fig. S4), a shift in $E_{\rm F}$ from +0.04 to +0.05 eV leads to a slight increase in $-\sigma_{zx}$, but a substantial decrease in $-\alpha_{zx}$ (Figs. 3e, 3f inset). This behaviour is consistent with the experimental observation, i.e., Mn doping suppresses $-\alpha_{zx}$ by 50 \% at low temperatures, while it enhances $-\sigma_{zx}$ only by $\sim 10$ \% (Figs. 3e, 3f). %\color{blue} In the low temperature limit, the transverse thermoelectric conductivity $\alpha_{ji}$ can be expressed as the energy derivative of the Hall conductivity (Supplementary Information). Thus, \color{red}  
	For a simple Weyl Hamiltonian, $\sigma_{ji}$ takes its maximum and $\alpha_{ji}$ becomes zero when $E_F$ is located exactly at the Weyl node and on the other hand, $\alpha_{ji}$ becomes strongly enhanced when $E_F$ moves slightly away\cite{Pallab2015}. Clearly, the real band structure of Mn$_3$Sn should be more complicated.  Interestingly, however, our theory also finds that $-\sigma_{zx}$ forms a peak at $E-E_{\rm F} \sim 0.065$ eV, around which Weyl nodes exist \cite{Yang2016}, and $-\alpha_{zx}$ becomes large when $E_F$ is slightly away from the nodes (Fig. S4).
%we can reproduce the trend in the temperature dependence of both $\sigma_{yx}$, and $\alpha_{yx}$ observed in experiment for Sample 1(2) (Figs. 3e, 3f inset).} 
%At $T >$ 100 K, the high-energy contribution dramatically enhances $\alpha_{yx}$. However, the high-energy states become irrelevant at low temperatures and a characteristic hump structure appears in the temperature dependence of $\alpha_{yx}$. 
%\new{Experimentally, we find that Mn doping suppresses $\alpha_yx$ by 50 \% at low temperatures, while it enhances $\sigma_{yx}$ only by $\sim 10$ \%. This is consistent with the theoretical expectation that the increase in $E_{\rm F}$ by Mn doping suppresses the Berry curvature at  $E_{\rm F}$ and thus leads to the formation of the peak in $\sigma_{yx}$, where the Weyl node exists\cite{Yang2016,Pallab2015}.}
 Our results thus indicate that the ANE in Mn$_3$Sn is particularly enhanced because of the characteristic structure of the Berry curvature with several Weyl points nearby the Fermi level\cite{Yang2016,Pallab2015}.  Our study further highlights the complementary roles of anomalous Hall conductivity $\sigma_{ji}$ and transverse thermoelectric conductivity $\alpha_{ji}$ in revealing the topological character of band structure. 

%In addition, recent theoretical calculation of the Nernst signal for a Weyl semimetal indicates that the large Berry curvature or fictitious field generated by the magnetic monopoles at the Weyl nodes not only produces a large Hall effect but a large Nernst signal when the Fermi level $E_{\rm F}$ is tuned close to the Weyl points\cite{Pallab2015}. According to the theory, a non-linear increase in the transverse thermoelectric conductivity and thus in the Nernst signal should arise as $E_{\rm F}$ shifts toward the Weyl points, following the Mott relation $\alpha_{yx} = \frac{\pi^2 k_{\rm B}^2 T}{3 e} (\frac{\partial\sigma_{\rm H}}{\partial{E}})_{E_{F}}$, where $k_{\rm B}$ is the Boltzmann constant\cite{Yong2008}.
% On the other hand, when the Weyl points locate exactly at $E_{\rm F}$, the ANE is expected to be small as the Hall effect becomes maximized with $(\frac{\partial\sigma_{\rm H}}{\partial{E}})_{E_{F}} = 0$. 
%Our observation of both large anomalous Nernst as well as Hall effects supports the idea that the Weyl points play a major role in their mechanism.
%The low temperature saturation of the Peltier conductivity may be consistent with the existence of the Weyl node. 
%In this case, a small tuning of the Fermi level, for example, by chemical substitution may well change the Nernst as well as Hall effects significantly. 

\color{black}
Finally, from the viewpoint of application for thermoelectric power generation, ANE could be useful as it facilitates the fabrication of a module structurally much simpler than the conventional one using the Seebeck effect\cite{Sakuraba2013}. The orthogonal orientation of the voltage output to the thermal heat flow (Fig. 1a) enables a lateral series connection of a single kind of ferromagnet with alternating magnetization direction (Fig. 1a inset). This simplifies a thermopile structure to efficiently cover the surface of a heat source (Fig. 1a inset), in comparison with the conventional thermoelectric module using the Seebeck effect, which consists of a pillar structure of alternating P- and N-type semiconductors. To increase power density, a thermopile should ideally cover the entire surface of a heat source, and therefore, a micro-fabricated thermopile array has to be arranged as densely as possible. However, as long as a ferromagnet is used, their inherent stray fields may perturb magnetization direction of neighboring modules, and limit the integration density. 

Our discovery of a new class of material that produces almost no stray fields but exhibits a large ANE is highly important for the application toward thermoelectric power generation, and it should allow us to design a thermopile with a much denser integration of thermoelectric modules to efficiently cover a heat source than the ferromagnetic counterparts.  While the observed values in this work would be still far from the size necessary for application, our study indicates that the magnetic Weyl metals such as Mn$_3$Sn would be particularly useful to obtain a large ANE by enhancing the Berry curvature at $E_{\rm F}$. Further studies to develop the technology for application such as thin film growth and coercivity control of such magnets will be important to build a thermoelectric power generator. 
%The observed ANE suggests that the power density using Mn$_3$Sn could reach $\sim 1 \mu$W/cc at room temperature. While the size is still far from the value necessary for application, the possible distinct mechanism based on the Weyl nodes in the $k$-space may well lead to a non-linear enhancement of the Nernst effect by Fermi level tuning, and would make it possible to build a thermoelectric power generator using an antiferromagnet in the near future.

\begin{figure}

		%\hspace{0cm} \includegraphics[width=2.5in]{fig1.eps} 
		\caption {{\bf Thermoelectric module using anomalous Nernst effect, and crystal and magnetic structures of Mn$_3$Sn.}  
			{\bf a,} Schematic illustration of a thermoelectric module based on the anomalous Nernst effect. 
			The anomalous Nernst electric field $\mbox{\boldmath $E$}$ appears in the direction of the outer product of the magnetization $\mbox{\boldmath $M$}$ and heat current $\mbox{\boldmath $Q$} \sim -\mbox{\boldmath $\nabla$} T$, and thus can be described as $\mbox{\boldmath $E$} = Q_{\rm s} \mu_0\mbox{\boldmath $M$} \times \mbox{\boldmath $-\nabla$}T$, where $Q_{\rm s}$ is the anomalous Nernst coefficient and $\mu_0$ is vacuum permeability. For the case of Mn$_3$Sn, the configuration of the sublattice moments is schematically presented.
			Inset: Schematic illustration of a thermopile made of an array of magnetic modules.  The in-plane magnetization directions of neighboring thermoelectric modules are alternated so that the Nernst signal with the same sign can be added up in series. The heat flows along the direction perpendicular to the basal plane of the heat source. %Small magnetization in an antiferromagnet produces almost no perturbing stray fields and allows the fabrication of a thermopile with high-density integration of thermoelectric modules to more efficiently cover the surface of a heat source than the ferromagnetic case. 
			{\bf b,} An individual $ab$-plane of Mn$_3$Sn. Spheres represent Mn (purple) and Sn (gray) atoms.  In addition to the unit cell frame, Mn atoms are connected by lines to illustrate that they form a breathing type of kagome lattice (alternating array of small and large triangles). Mn moments (arrow) form an inverse triangular spin structure. %Each Mn moment has the local easy-axis parallel to the in-plane direction toward its nearest neighbour Sn sites.
		} 

\end{figure}

\begin{figure}

%\hspace{-2cm} \includegraphics[width=6in]{fig2.eps} 
\caption {{\bf Magnetic field dependence of the anomalous Hall effect in Mn$_3$Sn.}
{\bf a \& b,} Field dependence of the Hall resistivity $\rho_{H}=-\rho_{zx}$ of {\bf a,} Sample 1 (Mn$_{3.06}$Sn$_{0.94}$), and {\bf b,} Sample 2 (Mn$_{3.09}$Sn$_{0.91}$), obtained at 100 K, and 300 K. 
{\bf c \& d,} Field dependence of the Hall conductivity $\sigma_{H}=-\sigma_{zx}$ of {\bf c,} Sample 1 and {\bf d,} Sample 2 obtained at 100 K, and 300 K.} 

\end{figure}

\begin{figure}

%\hspace{-2cm}
%\includegraphics[width=4in]{fig4.ps}
\caption { {\bf Anomalous Nernst effect in Mn$_3$Sn.} {\bf a,} Anisotropic field dependence of the Nernst signal $-S_{ji}$ of Sample 1. For comparison, the field dependence of the magnetization $M$ (right axis) for $B \parallel x$ is shown. {\bf b,} $-S_{zx}$ vs. $B$ of Samples 1 \& 2 measured at various temperatures. Here, Samples 1 \& 2 refer to crystals with Mn$_{3.06}$Sn$_{0.94}$, and  Mn$_{3.09}$Sn$_{0.91}$, respectively. 
{\bf c \& d,} $T$ dependence of the zero-field remnant Nernst signal $S_{ji}$ (solid symbol) obtained by the field sweep measurements of {\bf c,} Sample 1 and {\bf d,} Sample 2. The thermoelectric contribution estimated using the relation, $\rho_{jj} \alpha_{ji} = S_{ji} - S_{ii} (\rho_{ji}/\rho_{ii}$), is also shown (open symbol).
Insets: Seebeck coefficients $S_{xx}$, $S_{yy}$ vs. $T$ measured under zero field for {\bf c,} Sample 1 and {\bf d,} Sample 2. 
{\bf e,} Hall conductivity $-\sigma_{zx}$ vs. $T$ estimated using the relation $\sigma_{zx} = -\rho_{zx}/(\rho_{xx}\rho_{zz})$. The sudden drop in $-\sigma_{zx}$ for Sample 2 is due to the magnetic transition at $\sim 60$ K (see {\bf f}). Inset:  $-\sigma_{zx}$ vs. $T$ for $E - E_{F} =+40$ (blue) and +50 (red) meV obtained by the first-principles calculation.
{\bf f,} Transverse thermoelectric conductivity $-\alpha_{zx}$ vs. $T$ (left axis), estimated using the relation, $\alpha_{zx} = (S_{zx}/\rho_{zz})+\sigma_{zx}S_{xx}$ for the field sweep results (solid circle). For comparison, the $T$ dependences of $M$ (right axis) obtained in $B = 1000$ G $\parallel y$ using a field-cooling (FC, solid symbols) and zero-field-cooling (ZFC, open symbols) sequences are shown.
The FC and ZFC results bifurcate at $\sim$ 50 (60) K for Sample 1 (2) due to the magnetic transition.  Inset: $-\alpha_{zx}$ vs. $T$ for $E - E_{F} =$~+40 (blue) and +50 (red) meV obtained by the first-principles calculation. The error-bars are shown if they are larger than the symbol sizes and indicate the measurement errors that come from the uncertainties of their geometrical factors (Methods).}

\end{figure}

\begin{figure}

%\hspace{-2cm}
%\includegraphics[width=4in]{Figure1.eps}
\caption {\label{RT} {\bf Magnetization dependence of the spontaneous Nernst effect for ferromagnetic metals and Mn$_3$Sn.} Full logarithmic plot of the anomalous Nernst signal $|S_{ji}|$ vs. the magnetization $M$ for a variety of ferromagnetic metals and Mn$_3$Sn measured at various temperatures and fields (Methods). It shows the general trend for ferromagnets that $|S_{ji}|$ increases with $M$. The shaded region indicates the linear relation $|S_{ji}| = |Q_{\rm s}|\mu_0M$, with $|Q_{\rm s}|$ ranging from $\sim 0.05~\mu\mathrm{V}/\mathrm{KT}$ to $\sim 1~\mu\mathrm{V}/\mathrm{KT}$. The Nernst signal data points for Sample 1 (Mn$_{3.06}$Sn$_{0.94}$) obtained at various temperatures for $B \parallel $  $[01\bar{1}0]$  (blue closed circle) and for Sample 2 (Mn$_{3.09}$Sn$_{0.91}$)   for $B \parallel $  $[01\bar{1}0]$ (green closed circle) do not follow the relation, and reach almost the same value as the largest among ferromagnetic metals with three orders of magnitude smaller $M$.}

\end{figure}

	\pagebreak
	\begin{figure}
		\begin{center}
			\hspace{-1cm}
			\includegraphics[width=6in]{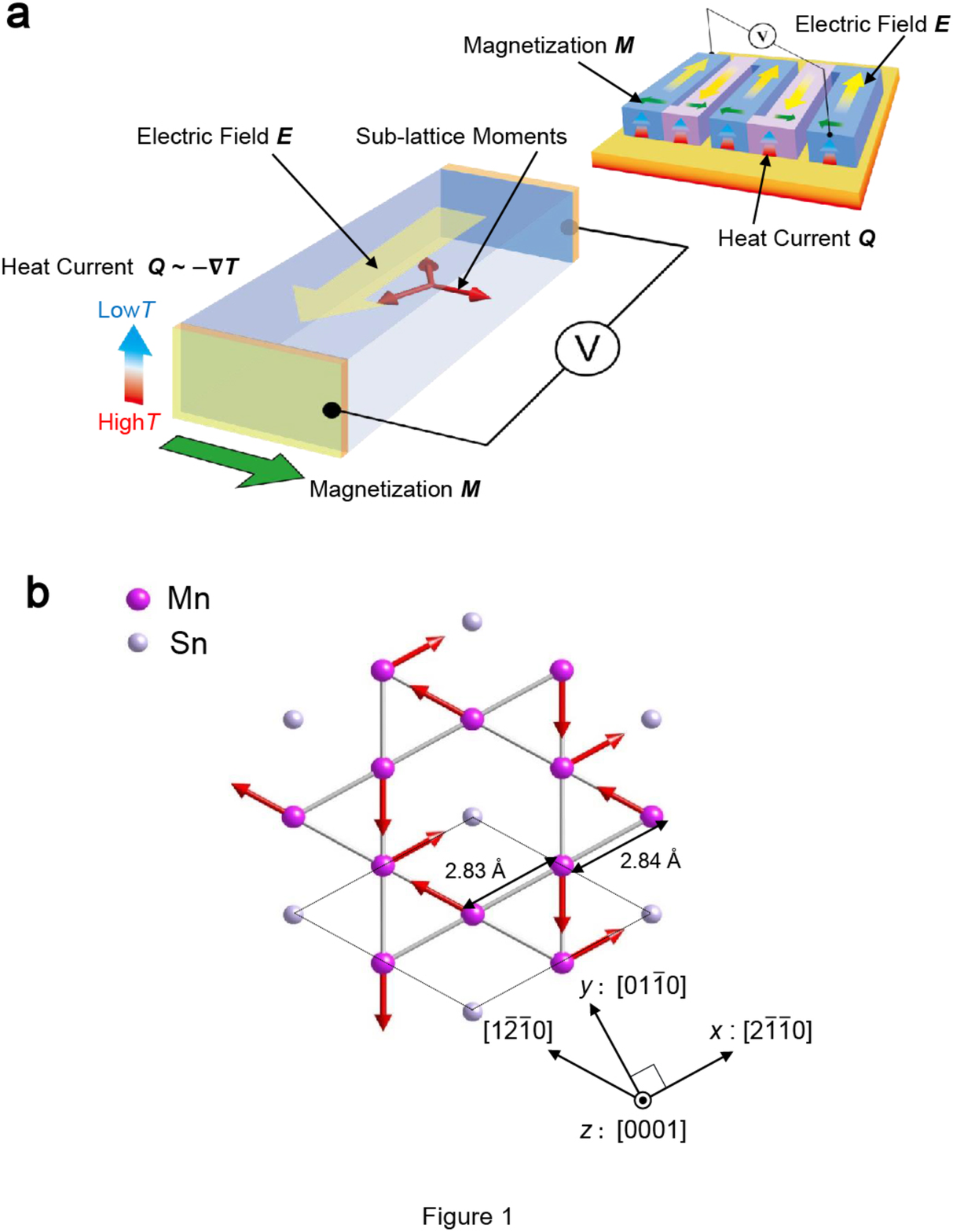}
		\end{center}
	\end{figure}

	\begin{figure}
		\begin{center}
			\hspace{-1cm}
			\includegraphics[width=6in]{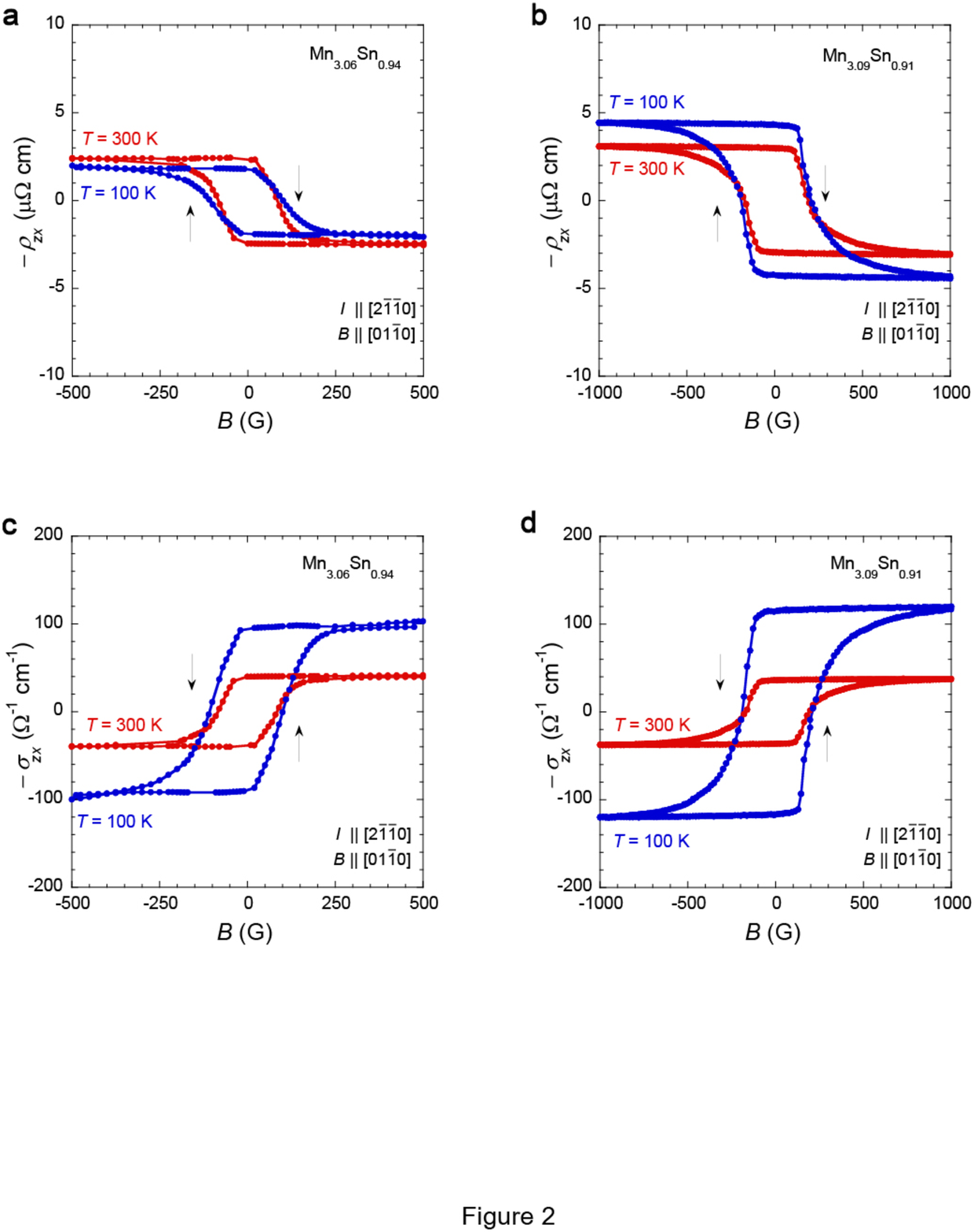}
		\end{center}
	\end{figure}

	\begin{figure}
		\begin{center}
			\hspace{-1cm}
			\includegraphics[width=6in]{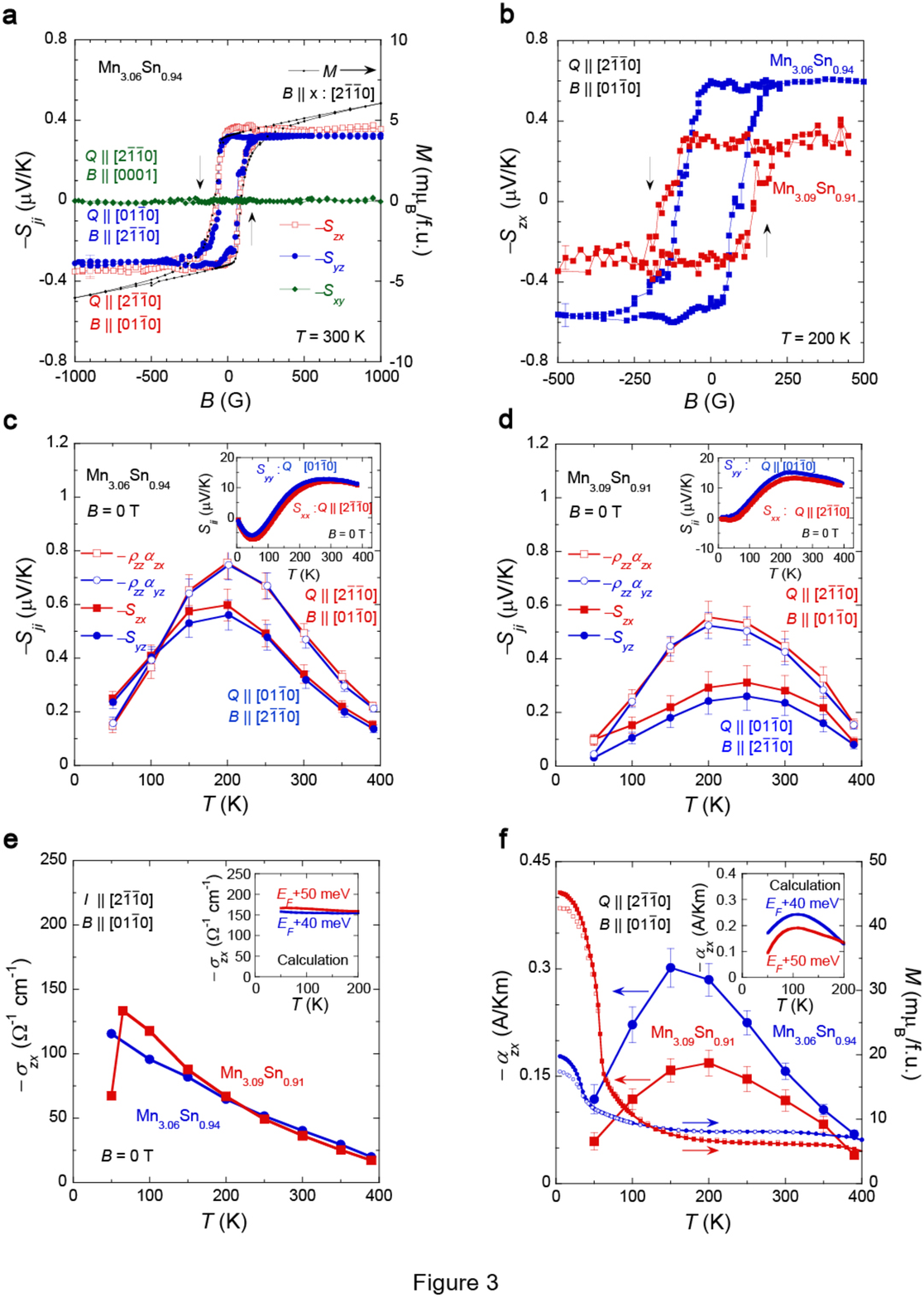}
		\end{center}
	\end{figure}

	\begin{figure}
		\begin{center}
			\hspace{-1cm}
			\includegraphics[width=6in]{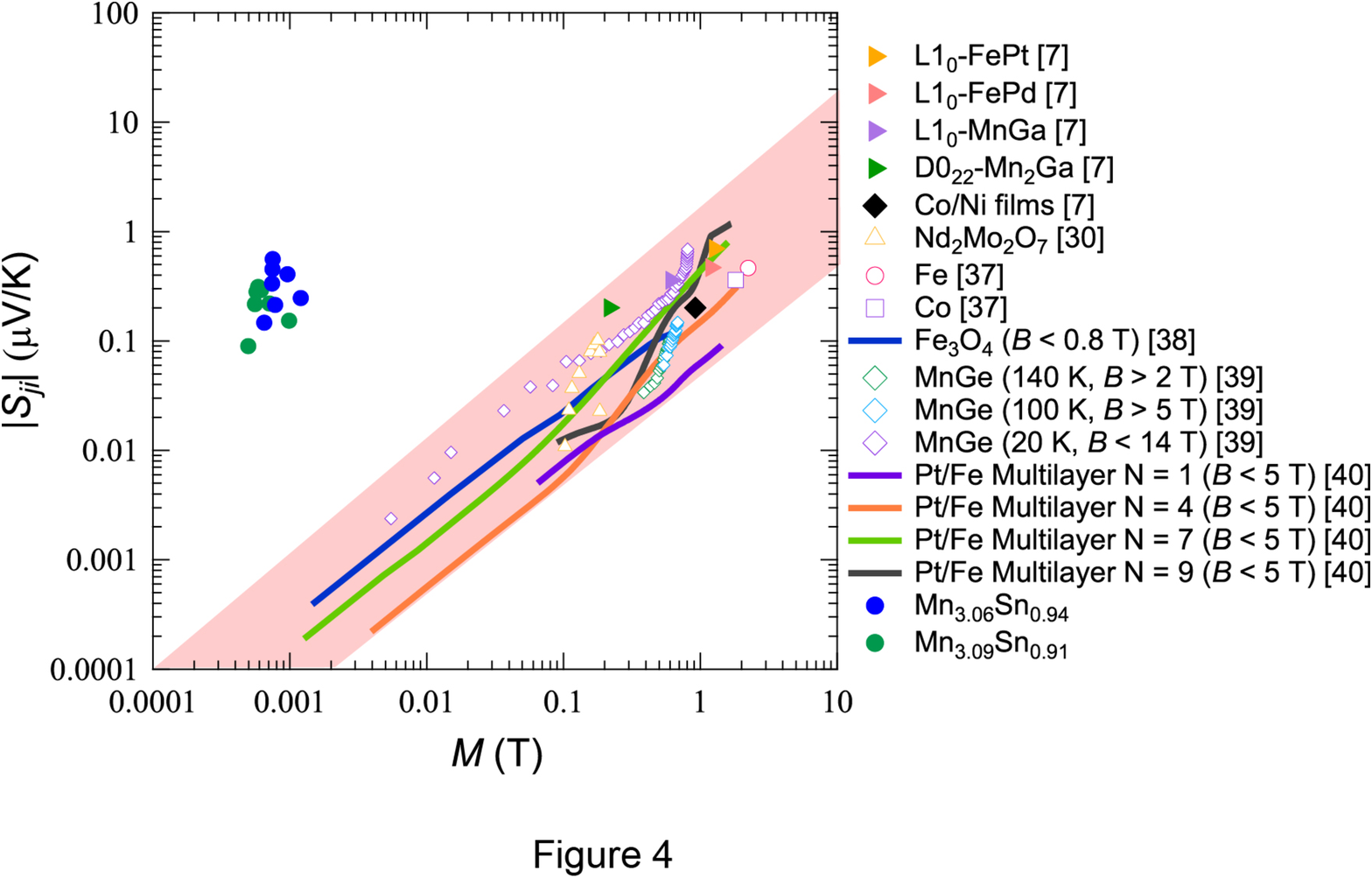}
		\end{center}
	\end{figure}
	\newpage	

\subsection{Methods}
Polycrystalline samples were made by melting the mixtures of manganese and tin in an alumina crucible sealed in an evacuated quartz ampoule in a box furnace at 1050 $^\circ$C for 6 hours. In preparation for single crystal growth, the obtained polycrystalline materials were crushed into powders, compacted into pellets, and inserted into an alumina crucible which was subsequently sealed in an evacuated silica ampoule. Single crystal growth was performed using a single-zone Bridgman furnace with a maximum temperature of 1080 $^\circ$C and growth speed of 1.5 mm/h. Analysis using an inductively coupled plasma (ICP) spectroscopy showed that the composition of the single crystal is Mn$_{3.06}$Sn$_{0.94}$ for Sample 1 and Mn$_{3.09}$Sn$_{0.91}$ for Sample 2. Powder X-ray diffraction measurement was performed using diffractometer (RAPID, Rigaku) with a rotating stage in order to remove preferred orientation in the data. One dimensional intensity pattern was extracted from two-dimensional Debye-Scherrer rings which were obtained using transmission mode. Rietveld analysis were performed using RIETAN-FP. All the samples were shown to be single phase, with lattice parameters consistent with previous work \color{black} (see Supplementary Information and Table S1\&S2).  After the alignment made using Laue diffractometer, the as-grown single crystals were cut into bar-shaped samples by spark machining for transport and magnetization measurements.

The magnetization measurements were done using a commercial SQUID magnetometer (MPMS, Quantum Design). The associated measurement error is less than a few \%. The specific heat measurement was performed using a commercial system (PPMS, Quantum Design) under zero magnetic field (Supplementary Information). Both longitudinal and Hall resistivities were measured by a standard
four-probe method using a commercial system (PPMS, Quantum Design). Electrical contacts were made by spot-welding gold wires for the longitudinal and transverse voltage probes, both of which were placed $\sim$ 1.5 mm apart on the sample with a typical current cross section area of $\sim$ 0.20 mm$^2$. In addition, thermoelectric properties were jointly measured by the one-heater and two-thermometer configuration using a commercial system (PPMS, Quantum Design). For thermoelectric measurements, the samples have the typical dimension of $\sim 10\times2\times2$ mm$^3$ for Sample 1 (Mn$_{3.06}$Sn$_{0.94}$) and $\sim5\times1.5\times1.5$ mm$^3$ for Sample 2 (Mn$_{3.09}$Sn$_{0.91}$). The thermal gradient $-\nabla T$ was applied by a heater at one end of the bar-shaped sample toward a thermal bath at the other end, and was measured by monitoring two thermometers linked to the sample by strips of $\sim 0.5$ mm-wide copper-gold plates along the longest direction of the sample. The distance between the two thermometers are approximately $\sim$ $5$ mm for Sample 1 and $\sim$ $2.5$ mm for Sample 2. The magnitude of the transverse voltage $\Delta V$ was found to be linearly increasing with the increase of applied temperature difference $\Delta T$. Here, $\Delta T$ was typically set to be 1.5 \% $\sim$ 2.0 \% of the sample temperature for both Seebeck and Nernst measurements. By setting the temperature gradient $-\nabla T$ along a bar-shaped single crystal ($x$-axis), the thermoelectric longitudinal and transverse emf voltages $V_{i}$ and $V_{j}$ were measured in an open circuit condition. The Seebeck coefficient $S_{ii}$ and Nernst signal $S_{ji}$ were then estimated as $S_{ii} = E_{i}/\nabla T$ and $S_{ji} = E_{j}/\nabla T$, where $E_i$ and $E_j$ are the longitudinal and transverse electric field. 
The magnetic field dependence of the Hall resistivity and the Nernst signal was obtained after removing the longitudinal component of the respective transport properties, which is found to be approximately constant as a function of the magnetic field (Supplementary Information).

The measurement errors for the longitudinal resistivity $\rho_{ii}$, Hall resistivity $\rho_{ji}$, and Nernst signal $S_{ji}$, are dominated by the uncertainties of their respective geometrical factors. The uncertainty of the longitudinal resistivity and Hall resistivity is 1-2\% and smaller than the symbol sizes used in Figures in the main text and Supplementary Information. The uncertainty in the Nernst signal $S_{ji}$ is $\sim 10\%$ for Sample 1 and $\sim 20\%$ for Sample 2. The corresponding error-bars for the Nernst signal and the transverse thermoelectric conductivity are given in Fig. 3 and Fig. S2.

The transverse thermoelectric conductivity $\alpha_{zx}$ was calculated using the Berry curvature formula\cite{Xiao2006} with the first-principles electronic structure. The density functional theory calculation was performed within the generalized-gradient approximation\cite{Perdew1996} as implemented in the quantum-ESPRESSO package\cite{giannozzi2009}. A 7$\times$7$\times$7 $k$-point grid, ultrasoft pseudopotentials\cite{vanderbilt1990} and plane wave basis sets with cutoff energies of 80 Ry for wave functions and 320 Ry for charge densities were used. For the Berry curvature calculation, a Wannier-interpolated band structure\cite{mostofi2008} with 40$\times$40$\times$40 $k$-point grid was employed.

The specimens for transmission electron microscope were prepared by Ar ion milling using
a JEOL Ion-Slicer operated at 5.5 kV and 150 $\mu$A under a low beam angle of 2.5$\degree$. Selected area
electron diffraction and high resolution lattice images were obtained using a transmission electron
microscope (JEOL JEM-2010F) operated at 200 kV. High resolution lattice images were simulated
by MacTempas software. 

Figure 4 was made using the ANE results obtained for various ferromagnets well below their Curie temperatures, as reported in literature including   L1$_0$-FePt  (300 K)\cite{Hasegawa2015}, L1$_0$-FePd  (300 K)\cite{Hasegawa2015}, L1$_0$-MnGa  (300 K)\cite{Hasegawa2015}, D0$_{22}$-Mn$_2$Ga  (300 K)\cite{Hasegawa2015}, Co/Ni films  (300 K)\cite{Hasegawa2015}, Nd$_2$Mo$_2$O$_7$  ($T<T_{c}= 73$ K, $B= 1$ T $\parallel$ [111])\cite{hanasaki2008}, Fe (300 K) \cite{weischenberg2013}, Co (300 K)\cite{weischenberg2013},  Fe$_3$O$_4$ (300 K, $B< 0.8$ T)\cite{ramos2014}, MnGe  (140 K, $B> 2$ T)\cite{Shiomi2013}, MnGe  (100 K, $B>5$ T)\cite{Shiomi2013}, MnGe  (20 K, $B<14$ T)\cite{Shiomi2013}, and  Pt/Fe Multilayer $N=1\sim9$ (300 K, $B<5$ T)\cite{Uchida2015}.
% The representation points for Pt/Fe multilayer are obtained by the linear interpolation between experimental data points.
\newpage

%Since the transverse magnetoresistance is relatively small in comparison with the Hall resistivity component,  

%Here is a description of a specific method used.  Note that the
%subsection heading ends with a full stop (period) and that the
%command is \verb|\subsection{}| not \verb|\subsection*{}|.

%\end{methods}
%% Put the bibliography here, most people will use BiBTeX in
%% which case the environment below should be replaced with
%% the \bibliography{} command.

\bibliography{Mn_Compound_ver31}

%% Here is the endmatter stuff: Supplementary Info, etc.
%% Use \item's to separate, default label is "Acknowledgements"

\begin{addendum}
 %\item[Supplementary Information] is linked to the online version of the paper at www.nature.com/nature.

 \item We thank Agung Nugroho, Akito Sakai, Tomoya Higo, Naoki Kiyohara for useful discussions. 
This work is partially supported by CREST and PRESTO, Japan Science and Technology Agency, and Grants-in-Aid for Scientific Research (16H02209), and Program for Advancing Strategic International Networks to Accelerate the Circulation of Talented Researchers (No. R2604) from
the Japanese Society for the Promotion of Science, and by Grants-in-Aids
for Scientific Research on Innovative Areas (15H05882, 15H05883, 26103002)
of the Ministry of Education, Culture, Sports, Science, and Technology of Japan.  The use of the facilities
of the Materials Design and Characterization Laboratory at the Institute for Solid State Physics, The University of Tokyo, is gratefully acknowledged.

 \item[Author Contributions] 
S.N. and Y.O. conceived the project. S.N. planned the experiments, and T.T., M.I., S.N. performed experiments and analyzed data. T.K., M.S., R.A. performed the first-principles calculations. D.N performed the TEM measurements and analyses. S.N., R.A. wrote the main text.  M.I., S.N., T.T., D.N, T.K., M.S., R.A. prepared the supplementary information and figures; All authors discussed the results and commented on the manuscript.

 \item[Competing Interests] The authors declare that they have no competing financial interests.

 \item[Correspondence] Correspondence and requests for materials should be addressed to S.N. \\(email: satoru@issp.u-tokyo.ac.jp).
 
 \item[Data Availability Statement]
The data that support the plots within this paper and other findings of this study are available from the corresponding author upon reasonable request
\end{addendum}

%\end{document}

\newpage
\title{\Large{\bf{Supplementary Information:\\ Observation of large anomalous Nernst effect at room temperature in a chiral antiferromagnet}}} 

%% Notice placement of commas and superscripts and use of &
%% in the author list

\author{Muhammad Ikhlas$^1$\footnote[1]{These two authors contributed equally.}, Takahiro Tomita$^1$\footnotemark[1], Takashi Koretsune$^{2,3}$, Michi-To Suzuki$^{2}$, Daisuke Nishio-Hamane\color{black}$^1$, Ryotaro Arita$^{2,4}$, Yoshichika Otani$^{1,2,4}$, Satoru Nakatsuji$^{1,4}$\\}

%\begin{document}
	
	\maketitle
	
	\begin{affiliations}
		\item Institute for Solid State Physics, University of Tokyo, Kashiwa 277-8581, Japan
		\item RIKEN-CEMS, 2-1 Hirosawa, Wako 351-0198, Japan		
		\item PRESTO, Japan Science and Technology Agency (JST), 4-1-8 Honcho Kawaguchi, Saitama 332-0012, Japan.		
		\item CREST, Japan Science and Technology Agency (JST), 4-1-8 Honcho Kawaguchi, Saitama 332-0012, Japan.		
		
	\end{affiliations}

	%While the Hall resistivity shows a significant jump as a function of magnetic field as shown in Fig. 2a in the main text, the longitudinal resistivity shows no change as a function of field at least for the field range $B \le 1$ T. A typical result of the magnetoresistance measured at various temperature is given in Supplementary Figure 1.
	
\section{Expression of the Hall conductivity $\sigma_{ji}$ and transverse thermoelectric conductivity $\alpha_{ji}$}

	To compare  the experimental results with first-principle calculation, we used the following expression for  the Hall conductivity $\sigma_{ji}$ and transverse thermoelectric conductivity $\alpha_{ji}$, which take into account the anisotropy of the longitudinal resistivity $\rho_{ii}$,
	\begin{eqnarray}
	\sigma_{ji}\approx-\frac{\rho_{ji}}{\rho_{ii}\rho_{jj}},\\ \alpha_{ji}\approx \frac{1}{\rho_{jj}}(S_{ji}-\frac{\rho_{ji}}{\rho_{ii}}S_{ii})
	\end{eqnarray}
	Here, ($i,j$) $=$ ($x,y$), ($y,z$), and ($z,x$), where $x$, $y$, and $z$ are taken to be the coordinates along [$2\bar{1}\bar{1}0$], [$01\bar{1}0$], and [$0001$], respectively \cite{Kiyohara2016}.  In our transport measurements, the applied current $I$ (heat current $Q = -\nabla T$) is defined to be parallel to the $i$-axis. The longitudinal resistivity $\rho_{ii}$ (Seebeck coefficient $S_{ii}$) is then measured using contacts placed along the $i$-axis while the Hall resistivity $\rho_{ji}$ (Nernst signal $S_{ji}$) is measured using contacts located along the $j$-axis. Using this definition, the sign of the Hall and the Nernst effect is determined by the right-hand rule.
	
	\color{black}

	\section{Temperature dependence of the longitudinal resistivity}
	The temperature dependence of the longitudinal resistivity for  Sample 1 (Mn$_{3.06}$Sn$_{0.94}$) and Sample 2 (Mn$_{3.09}$Sn$_{0.91}$) single crystals used is given in Figure S1. Both Samples 1$\&$2 show similar behaviour, i.e the in-plane resistivity is almost isotropic, while the out-of-plane resistivity shows a much stronger temperature dependence than the in-plane resistivity.   The residual resistivity ratio (RRR = $\rho$ (390 \ K) / $\rho$ (4.2 \ K)) which is a rough measure of the quality of the single crystals shows the value of RRR $\sim$ 1.5 -- 2.3 for Sample 1 and RRR $\sim$ 1.2 -- 1.35 for Sample 2 in the in-plane direction, and  $\sim$ 3.5  for Sample 1 and  $\sim$ 2.8 for Sample 2 in the out-of-plane direction. 
	This is consistent with the smaller coercivity observed in the Sample 1 crystals compared to Sample 2, as the coercivity in magnets generally increases with the number of defects and impurities, which may pin a magnetic domain wall. All the components of the longitudinal resistivity have continuous temperature change and nearly saturate above room temperature. \color{black} %\cut{All the components of the longitudinal resistivity either peak or saturate at around 300 K, indicating the existence of significant inelastic scattering. The $c$-axis component not only has smaller values at low temperatures than the in-plane counterparts, but shows stronger temperature dependence, indicating the band dispersion is stronger along the [0001] direction.} 
	
	%\section{Possible orbital contribution to the magnetization}
	%The Berry curvature or fictitious field in the $k$-space may also lead to the orbital magnetism due to the itinerant Hall current\cite{Shindou2001,Niu2010}. Thus, the existence of the large fictitious field equivalent to $\sim 1000$ T further indicates that there must be orbital moments in the material and the negative sign of the Hall conductivity suggests orbital ferromagnetism\cite{Shindou2001,Niu2010}. Indeed, the fact that the Nernst signal is proportional to the spontaneous part of the magnetization, but not to the field-induced part indicates that the spontaneous magnetization comes not only from the spin magnetization due to the spontaneous canting of the sub-lattice moments, but from the orbital moments induced by the large Berry curvature that produces both the large Hall and Nernst effects\cite{Xiao2006} (Fig. 3b in the main text). 
	
	\section{Magnetic field dependence of the Seebeck coefficient and the Nernst effect}
	The magnetic field dependence of the Seebeck coefficient $S_{ii}$ ($i= x, y$) was  measured at various temperatures under $B \ <$ 0.1 T along in-plane directions. We found no field dependence within experimental accuracy over all the temperature range between 50 K and 390 K. Figures S2a and S2b represent typical $S_{xx}$ vs. $B$   results obtained at 100 K and 300 K   for Samples 1$\&$2, respectively.
	The magnetic field dependences of the Nernst effect $S_{zx}$  under $B \ <$ 0.1 T along in-plane directions for Samples 1$\&$2 are shown in Figures S2c and S2d, respectively. 
	
	\section{Temperature dependence of the Seebeck effect}
	As shown in  Figure 3c inset in the main text, $S_{ii}(T)$ ($i = x, y$) of Sample 1 (Mn$_{3.06}$Sn$_{0.94}$) changes its sign below around 100 K and peaks with a negative amplitude at $\sim 50$ K. On the other hand, Figure 3d inset shows that a much weaker or nearly no sign change occurs for Sample 2 (Mn$_{3.09}$Sn$_{0.91}$). The Seebeck coefficient $S$ generally has two contributions: $S = S_d + S_g$. $S_d$ is the contribution from the charge carrier diffusion and $S_g$ is the contribution from the phonon drag effect due to additional charge carriers dragged by the phonon flow.
	%The $S(T)$ behaviour is determined by a balance between the diffusive charge carrier $S_g$ and the phonon-drag contributions $S_d$.
	The effect of the phonon drag typically causes a peak in the Seebeck coefficient at $\sim \Theta_{\rm D}/5$, where $\Theta_{\rm D}$ is the Debye temperature\cite{ziman1960}. We made a linear fit to the temperature dependence of the specific heat divided by temperature (Supplementary Figure S3) at the lowest temperature region (2 $-$4.5 K) using  the Einstein-Debye equation, $C/T=\gamma+\beta T^2$, where $\gamma$ and $\beta$ are the parameters for the electronic and lattice contributions to the specific heat, respectively. The fit gives $\beta=0.300 \pm 0.015$   mJ mol$^{-1}$ K$^{-4}$ corresponding to the Debye temperature of $\Theta_{\rm D}=280\pm 14$ K. As shown in Figure 3c inset of the main text, the minimum of the Seebeck coefficient is roughly located around $\Theta_{\rm D}/5 = 56$ K, suggesting that the low temperature behaviour may be related to the phonon drag effect.  The nearly no sign change in Sample 2 may be related to a larger amount of defects caused by doping, which reduces the phonon lifetime. \color{black}  However, further discussion on the temperature dependence of the Seebeck effect is beyond the scope of this paper, since the multiband nature of this system\cite{Yang2016} would necessitate more involved analysis. On the other hand, the observed Nernst effect is dominated by the anomalous (spontaneous) contribution and thus transverse thermoelectric conductivity $\alpha_{ji}$ mainly comes from the Berry curvature. Therefore, the contribution from phonon drag in $\alpha_{ji}$ is negligible.  %The sign of $S_g$ is the same sign as that of $S_d$ for the normal phonon-phonon scattering process, whereas the Umklapp process reverses the sign of $S_g$. The negative $S_g$ observed at low temperature indicates that the Umklapp process is the dominant scattering process in this temperature range. Thus, the sign change at 100 K is arisen by a crossover between the positive $S_g$ produced by hole carriers and the sign reversed (negative) phonon drag contribution $S_d$. 
	
	\section{ Crystal structure of  Mn$_3$Sn}
	RAPID (Rigaku) X-ray diffractometer (MoK$\alpha$, $\lambda$ = 0.7103 \AA) was used at room temperature to investigate the crystal structure of Mn$_3$Sn.  Two-dimensional Debye-Scherrer rings were obtained by the  X-ray intensity transmitted through the sample, which were then converted to one-dimensional intensity patterns.  The diffraction patterns were analyzed using the Rietveld analysis program RIETAN-FP to determine the precise crystal structure \cite{izumi2007} . We assign a hexagonal structure ($P6_3/mmc$) with the lattice parameters $a=$5.662 \AA~ and $c=$4.529 \AA \cite{tomiyoshi1982polarized}. Figure S5 shows the typical result of the X-ray powder pattern obtained for Sample 1 (Mn$_{3.06}$Sn$_{0.94}$). The atomic coordinates for the crystal parameter are shown in Table S1 for Sample 1 (Mn$_{3.06}$Sn$_{0.94}$) and Table S2 for Sample 2 (Mn$_{3.09}$Sn$_{0.91}$). Our results indicate that the samples used in our study have the same crystal structure as those used for previous neutron diffraction measurements\cite{tomiyoshi1982polarized,Brown1990}
	\color{black}.

	\begin{table*}[b]
		\caption{\label{tab:table1}  Crystal structure parameters refined by Rietveld analysis for Mn$_{3+0.06}$Sn$_{1-0.06}$ (Sample 1) with $P6_3/mmc$ structure at 300 K. The lattice parameters and the atomic positions of the Mn site are determined by the analysis, which is made using the X-ray diffraction spectra with MoK$\alpha$ radiation ($\lambda = 0.7103$ \AA). The final $R$ indicators are $R_{\rm WP}$=5.29, $R_{\rm e}$=9.45, and $S$=0.560 \cite{izumi2007}.}
		
		\begin{tabular}{lrllll}
			\multicolumn{2}{l}{Mn$_{3+0.06}$Sn$_{1-0.06}$} &\multicolumn{4}{c}{$V=125.747(8)$ \AA $^3$} \\
			\hline\hline
			\multicolumn{1}{l}{parameters (\AA)} &\multicolumn{1}{c}{} & \multicolumn{1}{c}{$a=5.6624(2)$} & \multicolumn{1}{c}{$b=5.6624(2)$} & \multicolumn{1}{c}{ $c=4.5286(2)$}&\multicolumn{1}{c}{} \\
			\hline
			\multicolumn{1}{l}{ Atom                    } &\multicolumn{1}{c}{Wyckoff position} & $x$ & $y$ &  $z$& Occupancy\\
			\multicolumn{1}{l}{ Mn                   } &\multicolumn{1}{c}{6h} &  0.8388(2) &  0.6777(3) &   1/4& 1\\
			\multicolumn{1}{l}{ Sn/Mn               } &\multicolumn{1}{c}{2c} &  1/3 &  2/3 &   1/4& (0.94/0.06)\\
			\hline\hline
		\end{tabular}
		\label{tab:Ret}
		
	\end{table*}
	
	\begin{table*}[b]
		\caption{\label{tab:table1}  Crystal structure parameters refined by Rietveld analysis for Mn$_{3+0.09}$Sn$_{1-0.09}$ (Sample 2) with $P6_3/mmc$ structure at 300 K. The lattice parameters and the atomic positions of the Mn site are determined by the analysis, which is made using the X-ray diffraction spectra with MoK$\alpha$ radiation ($\lambda = 0.7103$ \AA). The final $R$ indicators are $R_{\rm WP}$= 4.35, $R_{\rm e}$=13.0, and $S$=0.334 \cite{izumi2007}.}
		
		\begin{tabular}{lrllll}
			\multicolumn{2}{l}{Mn$_{3+0.09}$Sn$_{1-0.09}$} &\multicolumn{4}{c}{$125.623(9)$ \AA $^3$} \\
			\hline\hline
			\multicolumn{1}{l}{parameters  (\AA)} &\multicolumn{1}{c}{} & \multicolumn{1}{c}{$a=5.6587(2)$} & \multicolumn{1}{c}{$b=5.6587(2)$} & \multicolumn{1}{c}{ $c=4.5300(2)$}&\multicolumn{1}{c}{} \\
			\hline
			\multicolumn{1}{l}{ Atom                    } &\multicolumn{1}{c}{Wyckoff position} & $x$ & $y$ &  $z$& Occupancy\\
			\multicolumn{1}{l}{ Mn                   } &\multicolumn{1}{c}{6h} &  0.8402(2) &  0.6803(3) &   1/4& 1\\
			\multicolumn{1}{l}{ Sn/Mn               } &\multicolumn{1}{c}{2c} &  1/3 &  2/3 &   1/4& (0.91/0.09)\\
			\hline\hline
		\end{tabular}
		\label{tab:Ret}
		
	\end{table*}

\section{High-resolution transmission electron microscope (TEM) image of Mn$_{3}$Sn}   Figures S6a and S6b show the high resolution TEM image of Mn$_{3}$Sn (using Sample 1 as a representative) taken at room temperature for the $(0001)$ and $(2\bar{1}\bar{1}0)$ planes, respectively. Within the resolution permitted by the maximum operating voltage (200 kV) of the transmission electron microscope (JEOL JEM-2010F), the lattice parameters $a$ and $c$ obtained from the image are in good agreement with the results of our Rietveld analyses discussed above. The overlayed simulation images show the expected position of the Mn and Sn atoms within the lattice.

\section{Heat treatment effect on the temperature dependence of the magnetization}
	A recent paper reported the magnetization curves for Mn$_3$Sn at different temperatures and found that the in-plane coercive field disappears along with the spontaneous magnetization as the sample is cooled down below 270 K\cite{Duan2015}. In contrast to the temperature dependence of the magnetization of our crystal, an additional transition is observed below 270 K in both their magnetization and heat capacity data. They associated this with a transition from the triangular spin structure to a long-period helical spin configuration, which has been observed in earlier neutron scattering studies\cite{cable1993}. It has been known that this transition is observed most pronouncedly in single crystals that were annealed at temperatures below 800$^{\circ}$C\cite{ohmori1987}. Their single crystals were synthesized by slow-cooling the melt from 1000$^{\circ}$C to 600$^{\circ}$C, and thus a part of this slow-cooling process may play a role of the annealing procedure mentioned above. The mechanism behind this transition is still unknown. On the other hand, the single crystal used in our present paper was prepared from the Bridgman furnace without any additional annealing process, and thus exhibit no transition above 50 K\cite{tomiyoshi1986triangular}.
	
\section{Relation between the anomalous Nernst effect and the anomalous Hall conductivity}
	Both the intrinsic anomalous Hall conductivity $\sigma_{zx}$ and the intrinsic anomalous transverse thermoelectric conductivity $\alpha_{zx}$  are governed by the Berry curvature, ${\bm \Omega}_n(\bm k)$, as\cite{Xiao2006}
\begin{eqnarray}
	\sigma_{zx} &=& -\frac{e^2}{\hbar} \int \frac{d \bm k}{(2 \pi)^3} \Omega_{n,y}(\bm k) f_{n \bm k},\\
	\alpha_{zx} &=& -\frac{e}{T \hbar} \int \frac{d \bm k}{(2 \pi)^3} \Omega_{n,y}(\bm k) 
	\left\{ (\varepsilon_{n \bm k} - \mu) f_{n \bm k} + k_{\rm B} T \ln \left[ 1 + e^{-\beta (\varepsilon_{n \bm k} - \mu)} \right] \right\}.
\end{eqnarray}
	Here, $\varepsilon_{n \bm k}, f_{n \bm k}$ are the band energy and the Fermi-Dirac distribution function with the band index $n$ and the wave vector $\bm k$. Since $\left\{ (\varepsilon_{n \bm k} - \mu) f_{n \bm k} + k_{\rm B} T \ln \left[ 1 + e^{-\beta (\varepsilon_{n \bm k} - \mu)} \right] \right\}$
	is finite only around the Fermi energy, $\alpha_{zx}$ is determined by the Berry curvature around the Fermi energy whereas $\sigma_{zx}$ is the summation of all the Berry curvature below the Fermi energy.
	According to these equations, $\alpha_{zx}$ and $\sigma_{zx}$ are related as follows:
\begin{eqnarray}
	\alpha_{zx} &=& -\frac{1}{e} \int d \varepsilon \frac{\partial f}{\partial \mu} \sigma_{zx}(\varepsilon) \frac{\varepsilon - \mu}{T}.
\end{eqnarray}
	Thus, the size and the sign of $\alpha_{zx}$ are determined by the slope of $\sigma_{zx}(\varepsilon)$ as well as the Berry curvature around the Fermi energy.
	Indeed, at low temperatures, Eq.(5) can be approximated as
\begin{eqnarray}
		\alpha_{zx} &= \frac{\pi^2 k_B^2 T}{3 |e|}\frac{\partial \sigma_{zx}(\varepsilon)}{\partial \varepsilon} + O(T^3).
\end{eqnarray}
	Supplementary Figure S4 indicates the calculated result of the energy dependence of the anomalous Hall and transverse thermoelectric conductivity.
	According to the result, a slight shift of the Fermi energy leads to a change in the anomalous Hall conductivity.
	For example, at $\varepsilon = 0.04$ eV, $-\partial \sigma_{zx}/\partial \varepsilon \sim 1000$ $(\Omega {\rm cm})^{-1} (\rm {eV})^{-1}$.
	This means that $-\alpha_{zx} \sim 0.0024 T$ (A/Km) at low temperature limit, which is consistent with the value of $-\alpha_{zx} = 0.25$ (A/Km) at $T=100$ K in the inset of Fig. 3f.
	
	In our first-principles calculation for Mn$_{3}$Sn, the number of the Mn-s, Mn-d, Sn-s, Sn-p electrons are estimated to be 0.84, 5.85, 1.67, and 3.26, respectively.
	For Mn$_{3.06}$Sn$_{0.94}$, we can assume that the electron occupancy of the Sn-s,
	Sn-p and Mn-s orbital
	are the same as those of Mn$_{3}$Sn, since the low-energy states around the
	Fermi level are mainly formed
	by the Mn-d orbital. Then the number of electrons doped into the Mn-d
	orbital is estimated to be 0.024,
	which shifts the Fermi energy by 0.04 eV for Mn$_{3.06}$Sn$_{0.94}$. With a similar calculation, the
	shift of the Fermi energy
	is estimated to be 0.05 eV for Mn$_{3.09}$Sn$_{0.91}$.
	
	In experiment, a small change in the composition in the crystals used in the present study, namely Mn$_{3.06}$Sn$_{0.94}$ and  Mn$_{3.09}$Sn$_{0.91}$, may cause such a shift in $E_{\rm F}$. Although the Hall conductivity between the two crystals are almost the same at high temperatures,  the difference between the Hall conductivity of the two crystals becomes apparent at low temperatures, as shown in Fig. 3e in the main text. At 100 K,  Sample 2 (Mn$_{3.09}$Sn$_{0.91}$) shows larger anomalous Hall conductivity (lower anomalous transverse thermoelectric conductivity) than Sample 1  (Mn$_{3.06}$Sn$_{0.94}$), which semi-quantitatively agrees with first principle-calculation shown in Figure S4. Further investigation is necessary to make a more thorough, quantitative comparison between experiment and theory using samples with a variety of doping levels, and this defines a subject of future study.

	The DFT calculation does not take account of the temperature dependence of the local magnetic moment size. Generally, the states further apart from the Fermi energy are more affected by the change in the moment size. Thus, in comparison with the transverse thermoelectric conductivity, the theoretical estimate for the Hall conductivity is expected to have the larger deviation from experiment. This is because according to Eqs. (3) \& (4), the Hall effect is the sum of Berry curvature of all occupied bands, while the transverse thermoelectric conductivity only derives from the low energy states around $E_{F}$. In fact, we found that the disagreement between the theoretical and experimental values of the Hall conductivity is larger than the one for the transverse thermoelectric conductivity (Fig. 3 in the main text).

	\color{black}

\begin{figure}
	\begin{center}
		\hspace{-1cm}
		\includegraphics[width=6in]{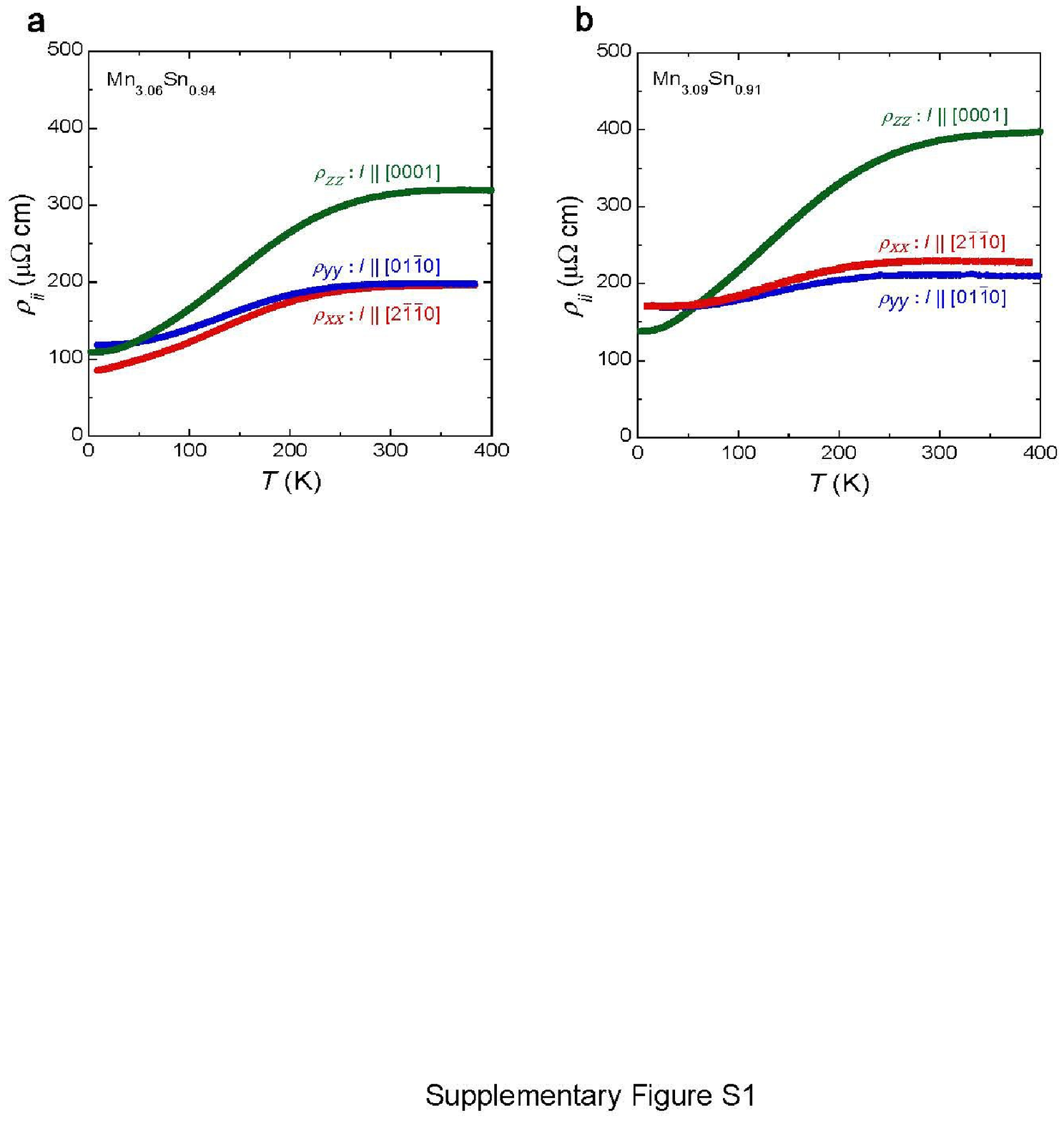}
	\end{center}
\end{figure}

\begin{figure}
	\begin{center}
		\hspace{-1cm}
		\includegraphics[width=6in]{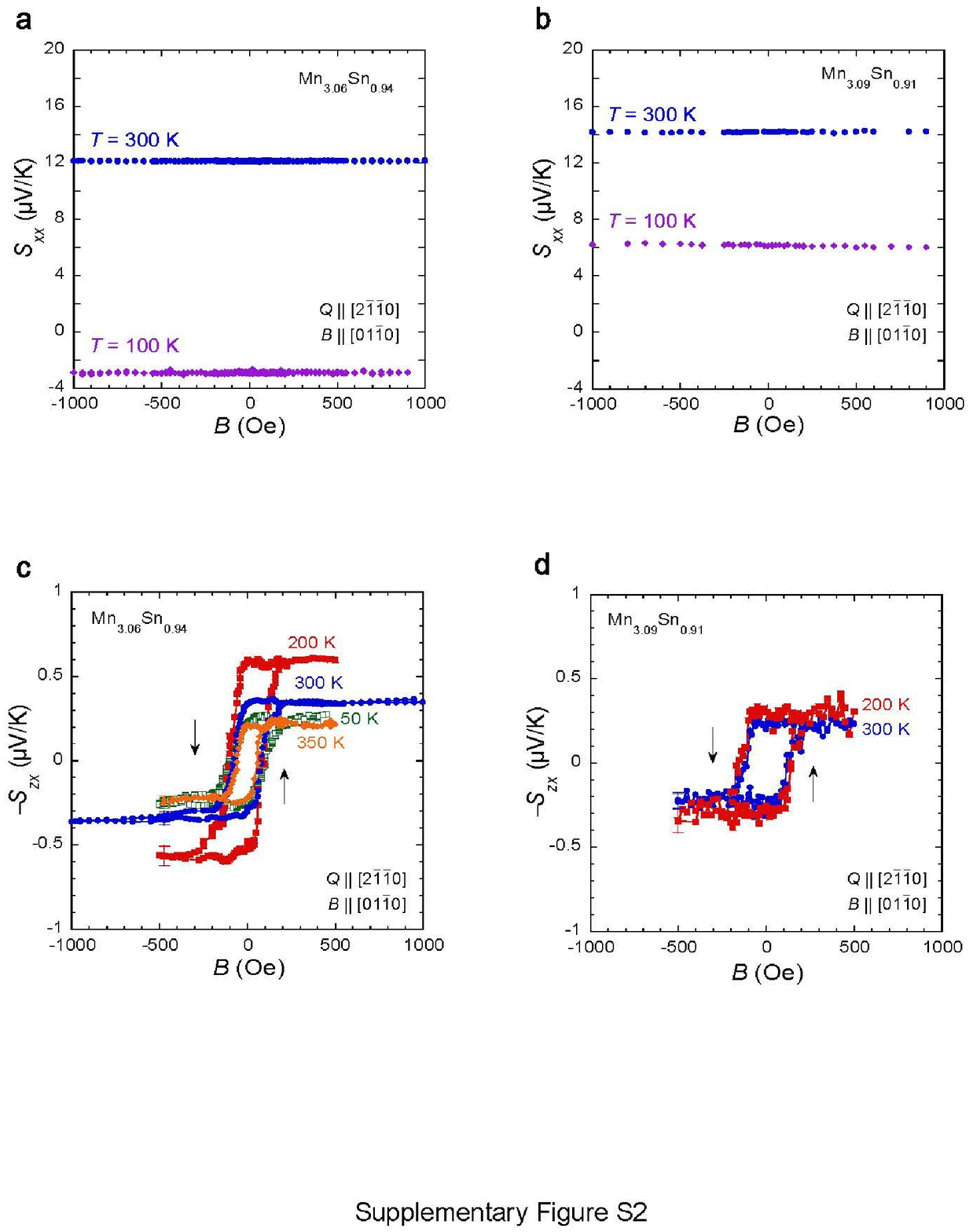}
	\end{center}
\end{figure}

\begin{figure}
	\begin{center}
		\hspace{-1cm}
		\includegraphics[width=6in]{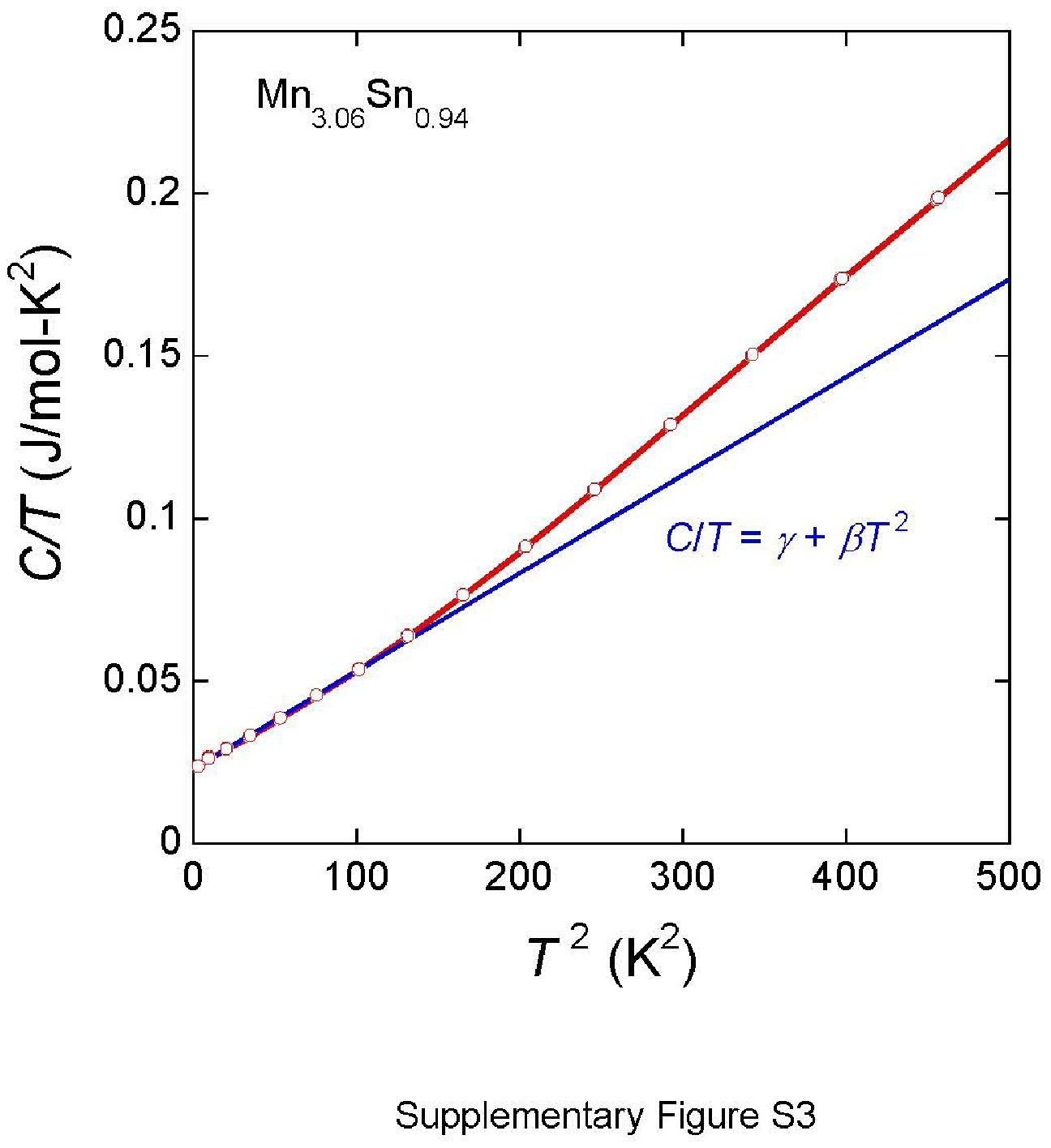}
	\end{center}
\end{figure}

\begin{figure}
	\begin{center}
		\hspace{-1cm}
		\includegraphics[width=6in]{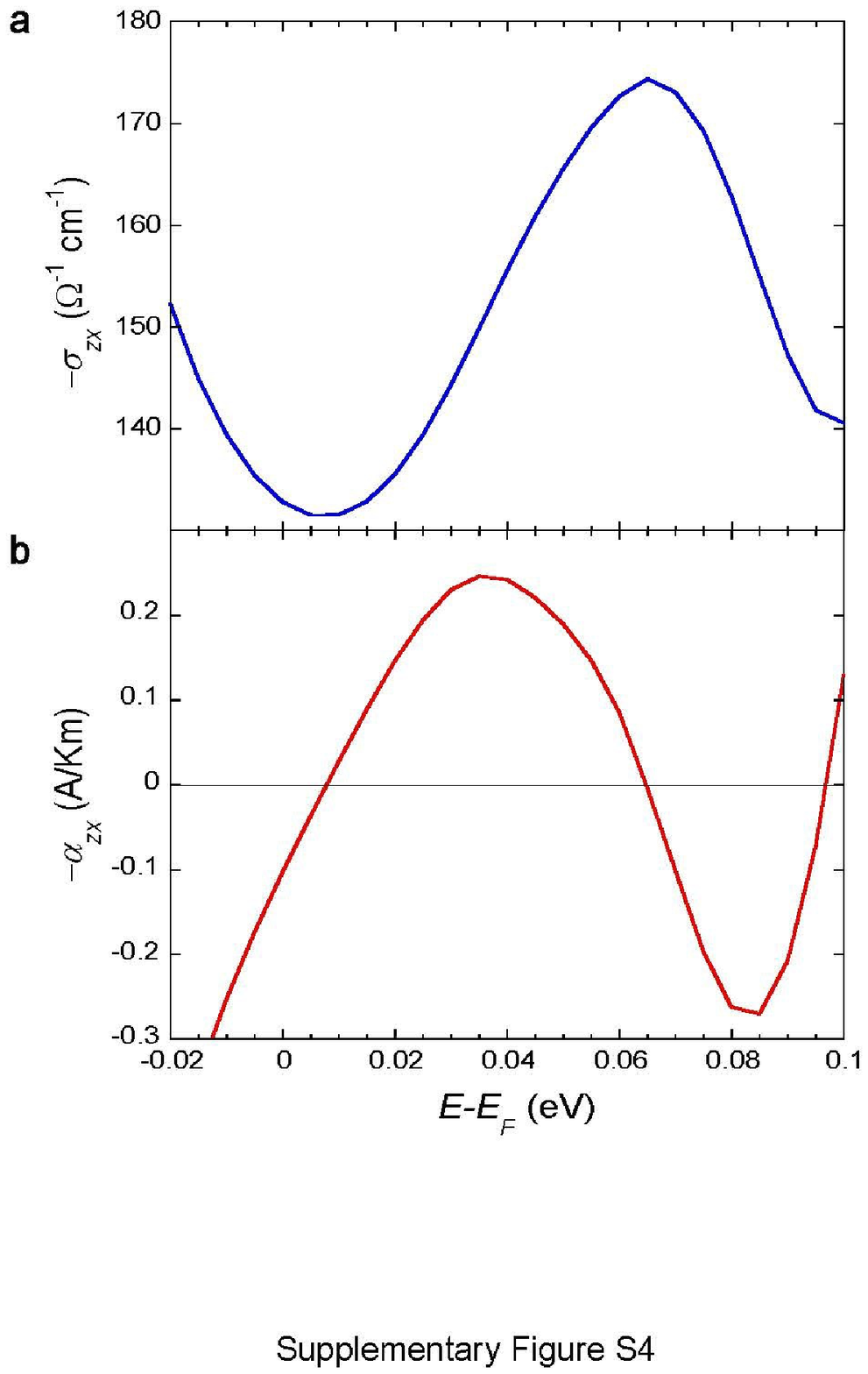}
	\end{center}
\end{figure}

\begin{figure}
	\begin{center}
		\hspace{-1cm}
		\includegraphics[width=6in]{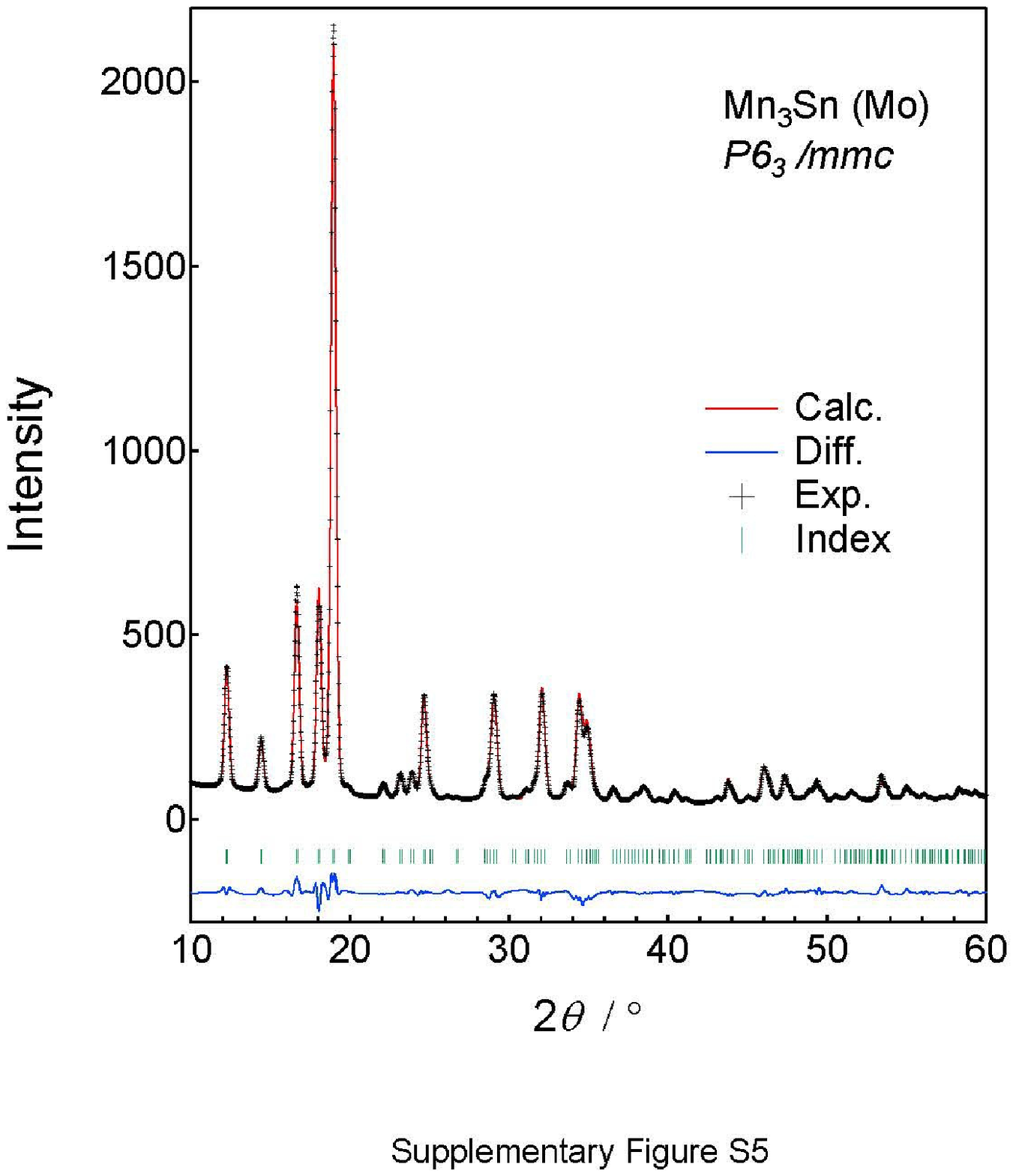}
	\end{center}
\end{figure}

\begin{figure}
	\begin{center}
		\hspace{-1cm}
		\includegraphics[width=6in]{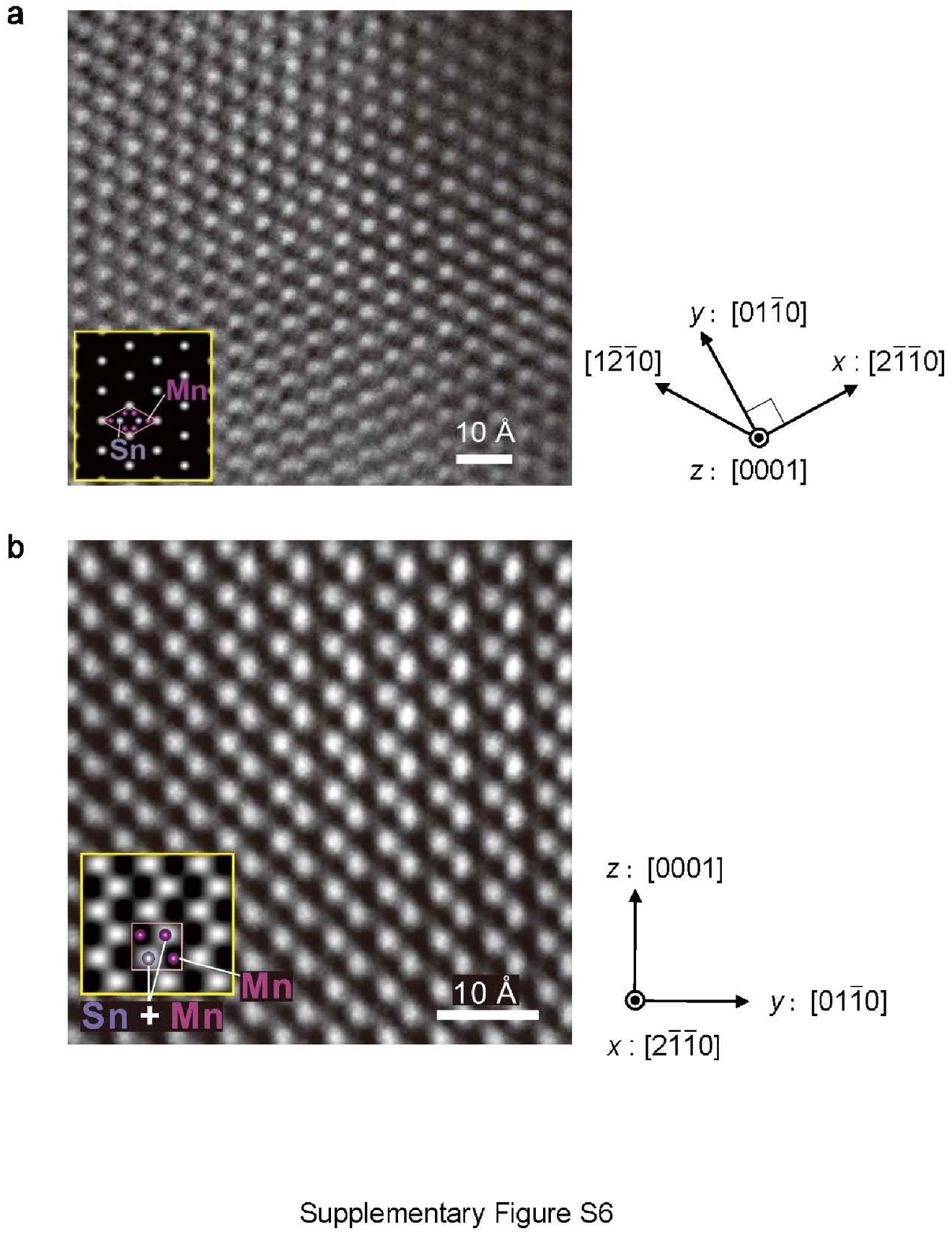}
	\end{center}
\end{figure}
\newpage	
	
	\begin{description}
		
		%	\item[Supplementary Figure 1.]
		%	{\bf Magnetic field dependence of the longitudinal resistivity}\\ Field dependence of the longitudinal resistivity $\rho(B)$ measured with the electric current $I\parallel [2\bar{1}\bar{1}0]$ at various temperatures.	
		\item[Supplementary Figure S1.]
		{\bf Temperature dependences of the longitudinal resistivity in Mn$_3$Sn}\\  Temperature dependence of the longitudinal resistivity $\rho_{ii}(T)$ ($i$ $=$ $x$ (red), $y$ (blue), and   $z$ (green)) of {\bf{a}}, Sample 1 (Mn$_{3.06}$Sn$_{0.94}$) and {\bf{b}},  Sample 2 (Mn$_{3.09}$Sn$_{0.91}$) measured at zero field.\color{black}
		
		\item[Supplementary Figure S2.]
		{\bf Magnetic field dependences of the Seebeck coefficient and Nernst effect in Mn$_3$Sn}\\  {\bf{a}}, Field dependence of the Seebeck coefficient $S_{ii}$ ($ i= x, y$) at 100 K and 300 K for Sample 1 (Mn$_{3.06}$Sn$_{0.94}$). {\bf{b}}, Field dependence of the Seebeck coefficient $S_{ii}$ ($ i= x, y$) at 100 K and 300 K for Sample 2 (Mn$_{3.09}$Sn$_{0.91}$). {\bf{c}}, Field dependence of the Nernst effect $-S_{zx}$ of Sample 1 (Mn$_{3.06}$Sn$_{0.94}$) at 350 K, 300 K, 200 K, and 50 K. {\bf{d}}, Field dependence of the Nernst effect $-S_{zx}$ of Sample 2 (Mn$_{3.09}$Sn$_{0.91}$) at 300 K and 200 K. The error-bars are shown if they are larger than the symbol sizes and indicate the measurement errors that come from the uncertainties of their geometrical factors (Methods)

		\item[Supplementary Figure S3.]
		{\bf Temperature dependence of the specific heat in Mn$_3$Sn}\\  Specific heat of  Sample 1 (Mn$_{3.06}$Sn$_{0.94}$) single crystal \color{black} measured in zero magnetic field divided by temperature $C/T$ (red circle), plotted vs. $T^2$. The blue solid line indicates a linear fit at the low temperature region, 2 $\le T \le$ 4.5 K, to estimate $\gamma$ and $\beta$ in the Einstein-Debye equation $C/T=\gamma+\beta T^2$. The fit gives the values $\gamma=22.7 \pm 0.2$ mJ mol$^{-1}$ K$^{-2}$ and $\beta=0.300 \pm 0.015$ mJ mol$^{-1}$ K$^{-4}$.
		
		\item[Supplementary Figure S4.]
		{\bf Energy dependence of the anomalous Hall conductivity and anomalous transverse thermoelectric conductivity}\\   Energy dependence of {\bf{a},} the anomalous Hall conductivity $-\sigma_{zx}$, and {\bf{b},}  the anomalous transverse thermoelectric conductivity $-\alpha_{zx}$ at $T=100$ K obtained by the first-principles calculation. $E_{\rm F}$ stands for the Fermi energy for the stoichiometric Mn$_3$Sn. %The black dotted line in both figures indicates the expected position of the Weyl nodes from the band structure calculation.  

		\item[Supplementary Figure S5.]
		{\bf X-ray diffraction pattern for Mn$_3$Sn}\\  Room temperature X-ray diffraction (XRD) patterns of Sample 1 Mn$_{3.06}$Sn$_{0.94}$. The crosses correspond to experimental data and the solid line (red) is for the Rietveld refinement fit. Vertical bars (green) below the curves indicate the peak positions of Mn$_3$Sn phase. The lower curve (blue) is the difference between the observed and calculated at each step.\color{black}
		
		\item[Supplementary Figure S6.]
		{\bf Transmission electron microscope image}\\ Bright field high resolution lattice image projected from  {\bf{a}}, [0001]  and {\bf{b}}, $[2\bar{1}\bar{1}0]$ directions, compared with simulation image. \color{black}
		
	\end{description}

\end{document}